\documentstyle[onecolumn]{mn}

\def\mincir{\raise -2.truept\hbox{\rlap{\hbox{$\sim$}}\raise5.truept
\hbox{$<$}\ }}
\def\magcir{\raise -2.truept\hbox{\rlap{\hbox{$\sim$}}\raise5.truept
\hbox{$>$}\ }}

\title{Physical constraints on the halo mass function}

\author[C. Porciani, F. Ferrini, F. Lucchin and
S. Matarrese]
{Cristiano Porciani$^{1,2}$,
Federico Ferrini$^{2}$,
Francesco Lucchin$^{3}$
and Sabino Matarrese$^{4}$ \\
$^{1}$ S.I.S.S.A. International School for Advanced Studies,
Strada Costiera 11, I--34014 Trieste, Italy\\
$^{2}$ Dipartimento di Fisica, Universit\`a di Pisa,
Piazza Torricelli 2, I--56100 Pisa, Italy\\
$^{3}$ Dipartimento di Astronomia, Universit\`{a} di Padova, vicolo
dell'Osservatorio 5, I--35122 Padova, Italy\\
$^{4}$ Dipartimento di Fisica {\em Galileo Galilei}, Universit\`{a} di
Padova, via Marzolo 8, I--35131 Padova, Italy\\}

\begin{document}

\maketitle

\begin{abstract}
We analyse the effect of two relevant physical constraints on the mass
multiplicity function of dark matter halos in a Press--Schechter type
algorithm.
Considering the random--walk of linear Gaussian density fluctuations as a
function of the smoothing scale, we simultaneously {\it i)} account for mass
semi--positivity and {\it ii)} avoid the {\em cloud--in--cloud} problem.
It is shown that the former constraint implies a severe cutoff of low--mass
objects, balanced by an increase on larger mass scales.
The analysis is performed both for scale--free power--spectra
 and for the standard
cold dark matter model. Our approach shows that the well--known
``infrared" divergence of the standard Press--Schechter
mass function is caused by unphysical, negative mass
events which inevitably occur in a Gaussian distribution of density
fluctuations.
\end{abstract}

\begin{keywords}
galaxies: clustering -- galaxies: formation --
large--scale structure of Universe
\end{keywords}

\section{Introduction}

The problem of understanding the origin and the evolution of the
density fluctuation field represents one of the fundamental issues of modern
cosmology. Primordial perturbations are believed to be generated from
vacuum fluctuations of a weakly coupled scalar field, 
during an inflationary stage. 
Because of this reason, it is often assumed that the
cosmological density fluctuations, during their linear evolution, made up a
Gaussian random field. This assumption might be also motivated on the basis
of the Central Limit Theorem.
It is well known, however, that this choice violates
the mass semi--positivity requirement ($\delta \ge -1$).
In fact, a Gaussian field with zero mean always admits a finite
probability of assigning events with values of the random variable
lower than $-1$; furthermore, this probability
raises when the field variance increases.
Thus, if one applies a coarse graining procedure to a hierarchical
Gaussian density field
(defined by $d\sigma ^2/dR_f <0$ and $\sigma ^2 \to +\infty$,
when the filtering length $R_f \to 0$),
one notes that the probability of finding regions with ``negative mass"
grows with decreasing $R_f$.
Therefore, we expect this feature of the Gaussian distribution to affect
the counting and arrangement properties of low--mass objects.

In this paper we study the effects caused by avoiding this shortcoming
of the Gaussian choice in obtaining the dark halo mass function
for a hierarchical scheme of structure formation.
Note that allowing for the physical constraint deriving from
mass semi--positivity amounts to introducing a sort of minimal
non--Gaussianity in the density fluctuation field. Some generic
features of the mass function obtained by non--Gaussian perturbations
have been already investigated by Lucchin \& Matarrese (1988), who found a
general tendency for an increase of the number of high--mass objects,
and by Sheth (1995).
A related problem has been studied by Catelan et al. (1994), who
analysed the effects of forbidding negative mass events on
the two--point correlation function of regions up--crossing a density
threshold.

We know that hierarchical clustering proceeds from
the ``bottom--up": low--mass halos form first while bigger
ones are created by aggregation and merging of still existing objects.
For this reason,
those models which aim at determining the mass function of cosmic structures
in a hierarchical aggregation scenario must afford the problem related
to the existence of sub--condensations inside clumps on larger scale.
This ``cloud--in--cloud" problem represents the main drawback of the classical
Press--Schechter theory (Press \& Schechter 1974)
in which, on the grounds of the spherical collapse model,
the virialized objects are identified with those regions
where the filtered density field, during its linear evolution,
becomes greater than a critical threshold, corresponding to a certain density
contrast $\delta_c$.

Our approach is an extension of the one developed by a number of authors
(Peacock \& Heavens 1990; Cole 1991; Bond et al. 1991), 
which succeeds in excluding sub--condensations from the count of bound
objects. [Alternative approaches to the mass function have been 
followed by Manrique \& Salvador--Sol\'e (1995) and Cavaliere et al. (1995).]
According to this method, in each point one considers the ``trajectory" 
$\delta(R_f)$ of the density field as a function of the 
filtering radius and determines the largest $R_f$ (and then the largest
possible mass) at which $\delta(R_f)$ crosses the threshold $\delta_c$.
This is enough to solve the cloud--in--cloud problem:
all the objects selected in this way cannot have been included in bigger
condensations, since the surrounding regions, filtered on all the larger scales
have density smaller than the critical one. Moreover, the objects which
could be associated to threshold crossings occurring on smaller
scales must not be considered, since the collapse of a structure erases
every sub--condensation. In much the same way we succeed in avoiding the
contribution from negative mass events to the count of dark matter halos
by imposing a boundary at $\delta_v\equiv -1$
to the random walk 
$\delta(R_f)$. This constraint has two important and complementary effects:
it implies a substantial decrease of the number of low--mass objects, thereby
eliminating the low--mass divergence of the Press--Schechter mass function,
which is balanced by an increase of the number of high--mass objects.
Moreover, also the redshift evolution of the resulting mass function is
largely modified compared to the standard Press--Schechter theory.

We mostly follow the formalism by Bond et al. (1991); in Section 2, we
derive the mass function according to their prescription while,
in Section 3 (but see also the appendix), we modify it to account
for the simple but physically relevant non--Gaussian feature
discussed above, which will be shown to have a strong impact on the
low--mass behaviour of the halo multiplicity function.
Section 4 contains a brief discussion and some conclusions. 

\section{The mass function of dark matter halos: a stochastic approach}

In this section we briefly review the mathematical formulation of the
``random--walk" approach sketched above.
After some general remarks, we focus on the configuration
that considers a sharp $k$--space filter in the
presence of a single absorbing barrier set at the threshold density.
We basically follow the approach by Bond et al. (1991),
although some aspects of the formulation are slightly modified.

Let us assume that the primordial density fluctuations
form a homogeneous and isotropic Gaussian random field
$\delta({\bf x},z)$, uniquely specified by its power--spectrum $P(k,z)$
(in the following, the power--spectrum at $z=0$ will be simply
denoted by $P(k)$).
If one identifies the collapsed regions with those points where
the filtered mass density field lies above a constant threshold, one can
allow for the redshift dependence of $\delta$ in terms of
the growing mode of linear perturbations, $D_+(z)$. Following
Bond et al. (1991), however, we ascribe the redshift dependence
to the formation threshold and consider the density fluctuations as a static
random field, $\delta({\bf x})$, normalized to its linear extrapolation to the
present time.
The evolution of $\delta_c$ is then fixed by the background cosmology:
the spherical collapse model gives $\delta_c(z)=\Delta (z)/D_+(z)$
where the function $\Delta (z)$ depends weakly on the values 
assumed at redshift $z$ by the density parameter
$\Omega$, the cosmological constant $\Lambda_{vac}$ and
the Hubble constant $H$ (Lilje 1992). In a matter dominated Einstein--de 
Sitter universe (the only case considered in this paper)
the threshold increases with redshift
according to the relation $\delta_c(z)=\delta_c/D_+(z)$, with $D_+(z) =
1/(1+z)$ and $\delta_c={\rm const}=1.686$.

By convolving the density field with the filter function
$W(|{\bf x}-{\bf x^\prime}|,R_f)$ one obtains a new field
$\delta({\bf x},R_f)$, which is defined in the four--dimensional space
$({\bf x}, R_f)$ and which, in general, is not homogeneous and isotropic
along the $R_f$ direction.
The dependence of $\delta({\bf x},R_f)$ on $R_f$ can be deduced with complete
generality using the Fourier transform of the density field,
$\delta({\bf x},R_f)=(2\pi)^{-3} \int \tilde \delta ({\bf k})
\widetilde W (kR_f) e^{-i{\bf k} \cdot {\bf x}} d^3k$.
In fact, an infinitesimal change of $R_f$ affects the value of
$\delta({\bf x},R_f)$ according to the relation
\begin{equation}
\label{lan}
{\partial \delta({\bf x},R_f) \over \partial R_f}=
{1 \over (2\pi)^3} \int \tilde \delta ({\bf k})
{\partial \widetilde W (kR_f) \over \partial R_f}
e^{-i{\bf k} \cdot {\bf x}} d^3k \equiv \eta({\bf x},R_f) .
\end{equation}

Due to the stochastic nature of $\delta({\bf x})$
and to the linearity of Eq. (\ref{lan}), it follows
that $\eta({\bf x},R_f)$ is also a zero mean Gaussian random field, which
is therefore uniquely defined through its auto--correlation function
\begin{equation}
\langle \eta({\bf x}_1,R_{f1}) \eta({\bf x}_2,R_{f2}) \rangle
= {1\over 2\pi^2} \int_0^\infty k_1^2 P(k_1)
{\partial \widetilde W (k_1R_{f1}) \over \partial R_{f1}}{\partial \widetilde W
(k_1R_{f2})
\over \partial R_{f2}}
{\sin (k_1r) \over k_1r} dk_1 ,
\end{equation}
obtained by using the definition of power--spectrum $\langle \tilde \delta
({\bf k}_1) \tilde \delta
({\bf k}_2) \rangle = (2\pi)^3 \delta_D ({\bf k}_1+{\bf k}_2) P(k_1)$,
where $\delta _D({\bf y})$ represents the Dirac delta function, 
and by defining $r=|{\bf x}_1-{\bf x}_2|$.

The equality that defines $\eta$ has the form of a Langevin equation for
the smoothed
density fluctuation field changing under the action of the stochastic force
$\eta({\bf x},R_f)$.
Unfortunately, for the most popular filter functions, such
as Gaussian and spherical top--hat, $\eta({\bf x},R_f)$ becomes a ``coloured"
noise, whose correlation properties along the $R_f$ direction make
the problem too involved to be afforded by analytical means.
It has been shown by Bond et al. (1991) that the problem becomes much more
tractable if one smooths the
density field by a ``sharp $k$--space'' filter, i.e. $\widetilde W
(k,k_f)=\theta(k_f-k)$,
where $\theta (x)$ is the Heaviside step--function and
$k_f \propto 1/R_f$. 
The calculation of the mean mass enclosed in the filtering volume has been 
performed by Lacey and Cole (1993) and gives $M(k_f)=6\pi^2 \langle 
\varrho \rangle /k_f^3$.
With such a filter, decreasing the radius corresponds
to adding up a new set of Fourier modes of the unsmoothed distribution to
$\delta(R_f)$; for a Gaussian random field, this
increment is completely independent of the previous steps,
so that the trajectory $\delta(R_f)$ represents a Brownian motion.

In practice, one can use the variable $k_f$ as ``time" variable, obtaining
\begin{equation}
{\partial \delta({\bf x},k_f) \over \partial k_f} =
{ 1 \over (2\pi)^3} \int \tilde \delta ({\bf k})
\delta_D (k_f-k) e^{-i{\bf k} \cdot {\bf x}} d^3k \equiv \xi({\bf x},k_f),
\end{equation}
where
 $\xi({\bf x},k_f)$ is a new Gaussian stochastic field.
By averaging over the statistical ensemble one then finds
$\langle \xi({\bf x},k_f) \rangle = 0$
and
\begin{equation}
 \langle \xi({\bf x}_1,k_{f1}) \xi({\bf x}_2,k_{f2}) \rangle
= {1\over 2\pi^2} k^2_{f1}P(k_{f1}){\sin (k_{f1}r) \over k_{f1}r}
\delta_D (k_{f1}-k_{f2}) .
\end{equation}

In an arbitrary point of space the density fluctuation field evolves with $k_f$
according to the Langevin equation
\begin{equation}\label{lpt}{\partial \delta(k_f) \over \partial k_f}=\xi(k_f) ,
\end{equation}
with the initial condition $\delta (0)=0$.
By averaging $\delta(k_f) =\int_0^{k_f}\!\!\xi(s)\, ds$ over the statistical
ensemble, one obtains
$\langle \delta(k_f) \rangle = 0$ and
$\langle \delta(k_{f1}) \delta(k_{f2}) \rangle =
(1/2\pi^2)\int _0 ^{\min (k_{f1},k_{f2})} s^2 P(s) \, ds$,
which completely determine the probability density
${\cal W}(\delta,k_f)$ of the Gaussian process $\delta(k_f)$, namely a
zero mean Gaussian distribution with 
variance $\sigma^2(k_f) = (1/ 2\pi^2) \int_0^{k_f} s^2 P(s) \, ds$.
These results show that our physical system is
dynamically equivalent to a set of particles undergoing one--dimensional
Brownian motion $x(t)$ with diffusion coefficient varying with time.
This analogy becomes even more evident if one identifies the
time variable with the variance $\Lambda \equiv \sigma^2(k_f)$ of the
filtered density field. In such a case, the stochastic process looses the
time--dependence of the diffusion coefficient and becomes a
Wiener process, $\partial \delta(\Lambda) / \partial \Lambda=\zeta(\Lambda)$,
with $\langle \zeta(\Lambda) \rangle=0$ and
$\langle \zeta(\Lambda_1) \zeta(\Lambda_2) \rangle =
\delta_D (\Lambda_1-\Lambda_2)$.

So far we have analysed the problem described by the Langevin
equation (\ref{lpt}) plus ``natural" boundary
conditions: $\lim_{\delta\to \pm \infty}
{\cal W}(\delta,k_f)=0$. Our aim is, however, to compute the fraction of
all trajectories which have crossed, at least once, the threshold for
structure formation at a given resolution $k_f$.
Such a quantity can be evaluated by putting an absorbing barrier
in $\delta=\delta_c(z)$: when a realization of the random process
$\delta (k_f)$ reaches for the first time the level $\delta_c$
one stops to count it, so that one always knows how many realizations
have not yet reached the barrier. In order to analytically solve this problem
it is more convenient to follow directly the behaviour
of the probability density ${\cal W}(\delta ,k_f)$, which can be easily shown
to obey the Fokker--Planck equation,
\begin{equation}
\label{fp}
{\partial {\cal W}(\delta ,\Lambda) \over \partial \Lambda} =
{1\over 2} {\partial ^2 {\cal W}(\delta, \Lambda)\over \partial \delta^2 } .
\end{equation}

The solution of equation (\ref{fp}) with absorbing boundary condition
in $\delta=\delta_c$, ${\cal W}(\delta_c ,\Lambda)=0$, and with the
initial condition ${\cal W}(\delta ,0)= \delta _D (\delta)$ has been obtained
for the first time by Chandrasekhar (1943); it reads
\begin{equation}
{\cal W}(\delta, \Lambda, \delta_c) =
{1\over \sqrt {2\pi \Lambda}} \left[ \exp \left(-{\delta ^2 \over 2\Lambda}
\right) - \exp \left(-{(\delta -2\delta_c)^2 \over 2\Lambda} \right)
\right] .
\end{equation}

By integrating the previous expression over the allowed region,
one obtains the probability that, by the ``time" $\Lambda$, a particle has
not yet
met the barrier, $S(\Lambda, \delta_c)= \int _{-\infty}^{\delta_c} \!{\cal
W}(\delta,
\Lambda, \delta_c) \,d\delta$.
Then the probability that, during its stochastic motion, a
given trajectory has crossed the critical level at a variance
lower than $\Lambda$
can be deduced from the probability conservation law,
$Q(\Lambda, \delta_c)= 1-S(\Lambda, \delta_c)$.
By differentiating with respect to $\Lambda$ one obtains
the probability distribution function of first--crossing variances,
\begin{equation} \label{tmp}
f(\Lambda,\delta_c)={\partial Q(\Lambda, \delta_c) \over \partial \Lambda} =
-{\partial  \over \partial \Lambda}\int_{-\infty}^{\delta_c}\! {\cal W}(\delta,
\Lambda, \delta_c) \,d\delta =
\left[- {1\over 2} {\partial {\cal W}(\delta,\Lambda, \delta_c)
\over \partial \delta}
\right] _{-\infty}^{\delta_c}=
{\delta_c \over \sqrt {2\pi} \Lambda^{3/2} }
\exp \left( -{\delta_c^2 \over 2 \Lambda} \right) .
\end{equation}

Bond et al. (1991) used these results to get an improved,
Press--Schechter--like expression for the mass function free of the
cloud--in--cloud problem.
The function $f(\Lambda,\delta_c) \,d\Lambda$ yields the
probability that a realization of the random walk is absorbed
by the barrier during the time interval $(\Lambda, \Lambda+d\Lambda)$,
or, thanks to the ergodic theorem, the probability that a point
is involved in the collapse of a structure in the mass
range [$M(\Lambda+d\Lambda), M(\Lambda)$].
The comoving number density of structures with mass in the range $(M, M+ dM)$
present at redshift $z$ is therefore given by
\begin{equation}
\label{genps}
n(M,z) dM = {\langle \varrho \rangle \over M} f(\Lambda,\delta_c(z))
\left| {d\Lambda \over dM} \right| dM ,
\end{equation}
which, using equation (\ref{tmp}), becomes
\begin{equation}
\label{ps}
n(M,z) d M =
{\langle \varrho \rangle \delta_c (1+z) \over \sqrt {2\pi}}
{1\over M^2 \Lambda ^{1/2} (M)} \left| {d\ln \Lambda \over d\ln M} \right|
\exp \left( - {\delta_c^2(1+ z)^2 \over 2 \Lambda (M)} \right) d M.
\end{equation}
This equation is identical to the Press--Schechter
formula, including the well--known ``fudge factor" of two.

\section{The zero mass barrier}

The technique adopted to solve the cloud--in--cloud problem
hides an inconsistency that is peculiar to every model which tries to
describe the density fluctuations by means of a Gaussian field. Indeed,
to deal with a universe which does not contain regions of negative mass,
one must assume that the density field
$\varrho({\bf x})$ is semi--positive definite and, as a consequence,
that the corresponding fluctuations $\delta({\bf x})$
get only values larger than or equal to $-1$ everywhere in space.
Moreover, if one considers
a window function that is semi--positive definite 
(except for a set of measure zero), also the filtered
density fluctuation field must
only get values $\delta (R_f)\geq \delta_v \equiv -1$.
All this conflicts with the Gaussian hypothesis which
in any point predicts $\delta (R_f)< -1$ with finite probability.

We want now to modify the algorithm presented in the previous section to
account for this physical constraint. Unfortunately one cannot easily forsake
the Gaussian assumption, because of its relevant role in the construction of
the Fokker--Planck equation ({\em Pawula theorem}; see, e.g., Risken 1989),
so one should devise a stratagem able to provide a theory equivalent to a
non--Gaussian one, but which remains tractable.

The existence of regions with 
negative mass ($\delta < -1$) in the Press--Schechter approach 
is caused by two main reasons: {\it i)} the initial conditions are represented 
by a Gaussian field; {\it ii)} the perturbations are evolved 
within the linear approximation. 
While the first motivation represents a real inconsistency if applied to the 
density fluctuation field, the second one
is much more subtle: we know that the real evolution of the perturbation
field differs sensibly from its linear approximation. Following the 
exact dynamics and starting from a self--consistent distribution of the
density field (i.e. one not containing negative mass events) one would never 
obtain regions with negative mass. 
However, in order to make quantitative predictions, one extrapolates
the validity of the linear theory beyond its limits and, keeping in mind
the spherical model, one assigns a rule--of--thumb for modelling the collapse 
of overdense fluctuations. Similarly, one should remind that underdense 
fluctuations also experience non--linear evolution, so that one could devise 
a suitable short--cut to account for the non--linear dynamics of underdense 
regions. In this section we treat this problem in a formal way:
we explicitly forbid our random trajectories to enter the unphysical region, 
by putting a reflecting barrier in $\delta=\delta_v\equiv-1$. 
To account for the time dependence of the density
fluctuation field, we assume that the value $\delta_v$ increases
with redshift according to $\delta_v(z) = \delta_v/D_+ (z)$.
The aim of this approach is to understand which features of
the mass--function actually depend on the presence of negative mass events 
in the probability distribution. 

The possibility to obtain an analytical solution is once again restricted to
the choice of the sharp $k$--space filter, which is, however, slightly
inconsistent with the semi--positivity assumption for the filter function.
In fact, the sharp cut in Fourier space implies, by the indeterminacy
principle, an infinite series of oscillations in configuration space where
the filter assumes many times negative values. On the other hand,
the absolute value of this function decreases quite rapidly as one goes away
from the origin, so that the integral which defines the filtered density field
is largely dominated by regions where the filter is positive.
We may then expect that the error one makes by using this window function
is small (notice, moreover, that a similar problem is present for the absorbing
barrier set at $\delta_c$).

The choice of a sharp $k$--space filter together with the presence of
a reflecting barrier in $\delta=\delta_v$ and of an absorbing one in
$\delta=\delta_c$ allows to write the
Fokker--Planck equation (\ref{fp}) for the probability density
${\cal W}(\delta,\Lambda)$,
with the boundary conditions
\begin{equation}
{\cal J}(\delta_v,\Lambda)=-{1\over 2}
{\partial {\cal W}(\delta,\Lambda) \over \partial \delta} \bigg|_{\delta_v}=0 ,
\ \ \ \ \ \ \ \ \ \ \ {\cal W}(\delta_c,\Lambda)=0
\end{equation}
and the initial condition ${\cal W}(\delta,0)=\delta_D(\delta)$.
Here ${\cal J}(\delta,\Lambda)$ represents the probability density current
and the equation which involves it characterizes the reflecting barrier.

In this case the evolution of the system is analogous to that of a set of
particles undergoing Brownian motion in the presence of
the potential $V(x)= +\infty$ if $x\le \delta_v$,
$V(x)= {\rm const}$ if $\delta_v<x<\delta_c$ and
$V(x) = -\infty$ if $x \ge \delta_c$.
Starting from this analogy, one may wonder for which properties of
the first--crossing distribution one should expect relevant
differences from the case previously considered.
Indeed, it is easy to deduce that the first--crossing times
are smaller on the average, since the reflecting barrier reverses
the motion of those particles which hit it, forbidding
their dispersion to very large distances away from the absorbing boundary.
As in our analogy the time variable corresponds to the variance of the
filtered density field, one should expect that the reflecting barrier
increases the number of crossings at small variance;
in practice, the numerical density of small mass objects should decrease while
that of large mass clumps should increase.
On the contrary, those realizations that reach the critical level in a very
short time, describing a ``quasi--coherent'' trajectory headed
forward, are not influenced by the presence of the reflecting barrier.
One should then expect the numerical density of very large mass objects
to be unaffected by our procedure.
A numerical computation of the mass function performed by using the adhesion
approximation in one dimension seems to share the same behaviour (Williams 
et al. 1991). However, we stress that N--body simulations of hierarchical 
clustering in three dimensions show better agreement with 
Press--Schechter results (Lacey \& Cole 1994). 

To solve the Fokker--Planck equation let us call $x$
the random field $\delta$ and
$t$ the independent variable $\Lambda$. Call then $A_c$ the position of the
absorbing barrier and $-R_v$ that of the reflecting one.
As initial condition one assumes a Dirac delta function set in $x=0$.
To simplify the equations one can shift the origin so that it corresponds to
the location of the reflecting boundary; in such a way the absorbing barrier
is found in $x=A_c+R_v$ and all Brownian particles leave from $x=R_v$.
One has then to solve the equation
\begin{equation}
\label{fpx}
{\partial {\cal W}(x,t) \over \partial t}={1\over 2}
{\partial^2 {\cal W}(x,t) \over \partial x^2}
\end{equation}
together with the boundary conditions
$\partial {\cal W}(0,t) / \partial x = 0$,
${\cal W}(A_c+R_v,t)=0$ and the initial condition
${\cal W}(x,0)=\delta_D(x-R_v)$.
The problem can be solved by separation of variables. One finds
${\cal W}(x,t) = 2 \sum_{n=0}^\infty \phi_n(x) \phi_n(R_v) \exp(-{1\over 2}
\lambda_n t)$, with eigenvalues $\lambda_n= [(2n+1)\pi/2(A_c+R_v)]^2$ and
orthonormal eigenfunctions $\phi_n(x) = (A_c+R_v)^{-1/2}
\cos\left(\sqrt{\lambda_n} x\right)$.
The probability density function then reads
\begin{equation}
{\cal W}(x,t)={2\over A_c+R_v} \sum _{n=0}^\infty
\cos\left({(2n+1)\pi\over 2(A_c+R_v)}R_v\right) \cos\left({(2n+1)\pi\over
2(A_c+R_v)}x\right)
\exp\left(-{(2n+1)^2\pi^2\over 8(A_c+R_v)^2}t\right) ,
\end{equation}
while the first--crossing rate across the absorbing barrier in $x=A_c$ is
\begin{equation}
{\cal T}_c(t)={\pi\over 2(A_c+R_v)^2} \sum_{n=0}^\infty
(-1)^n (2n+1) \cos\left({(2n+1)\pi\over 2(A_c+R_v)}R_v\right)
\exp\left(-{(2n+1)^2\pi^2\over 8(A_c+R_v)^2}t\right) .
\end{equation}

To obtain the mass function deriving from the above crossing rate one
must go back to the original physical variables
$\delta$ and $\Lambda$ and replace $f(\Lambda,\delta_c(z))\equiv
{\cal T}_c(\Lambda)$ in Eq.(\ref{genps}). Assuming $\Omega =1$ one gets
\begin{equation}
\label {rps}
n(M,z)={\pi \langle \varrho \rangle
\over 2\left(1+\delta_c\right)^2 (1+z)^2}{\Lambda (M)\over
M^2}\left|{d\ln \Lambda \over d\ln M}
\right| \sum_{n=0}^\infty (-1)^n (2n+1)
\cos\left( {(2n+1)\pi\over 2(1+\delta_c)
}\right)
 \exp \left( -{(2n+1)^2 \pi^2 \Lambda (M) \over
8 (1+\delta_c)^2(1+z)^2} \right) .
\end{equation}

We can now compare our result with the Press--Schechter mass function
for scale--free power--spectra $P(k)=A k^{n_p}$, where $n_p$ is the primordial
spectral index (let us remind that detailed analyses of the mass function
of clumps in $N$--body simulations with scale--free initial conditions have
been performed by Efstathiou et al. 1988 and by Lacey and Cole 1994).
In such a case, selecting a sharp $k$--space filter, one has
$\Lambda(M)=\sigma^2(M,z=0)= (M/ M_0)^{-2\alpha}$ with
$M_0\equiv 6\pi^2\langle \varrho \rangle \left(A / 12\pi^2 \alpha
\right)^{1/2\alpha}$
and $\alpha\equiv (n_p+3)/6$.
Replacing these expressions in Eq.(\ref{rps}), we
obtain
\begin {equation}
\label {mr}
n(M,z)={8 \alpha \over \pi}{ \langle \varrho  \rangle \over M^2}
\left( {M\over M_R(z)} \right) ^{-2\alpha}
\sum_{n=0}^\infty (-1)^n (2n+1)
\cos\left( {(2n+1)\pi\over 2(1+\delta_c)
}\right) \exp \left( -{(2n+1)^2
\left({M\over M_R(z)}\right) ^{-2\alpha}}   \right) ,
\end{equation}
where $M_R(z) \equiv M_0 \left[ \pi^2/ 8 (1+\delta_c)^2(1+z)^2
\right] ^{1/2\alpha}$
represents the characteristic mass of the distribution at redshift z.
Since the function $M^2n(M,z)$ depends only on the ratio $M/M_R(z)$,
our solution keeps the self--similarity property characterizing the
Press--Schechter theory.

In Figure 1 we show the behaviour of our new mass function
by plotting $M_R^2(z) n(M,z) /\langle \varrho \rangle$ which
depends only on the ratio $M/M_R$ and is not affected by
the normalization of the power--spectrum. 
We consider different values of the spectral index
($n_p=-2,~-1,~0,~1$) comparing our solution with the
Press--Schechter one. It is evident
that these mass functions are very different in the low--mass tail
while the discrepancy tends to disappear for large masses.
In fact $n(M)$ defined in Eq.(\ref{ps}) diverges as $M\to 0$
whereas ours goes to zero in the same limit.

This behaviour is quite interesting: in fact it is rather unnatural to
imagine that $n(M)$ grows unbounded for
$M \to 0$ in a hierarchical scenario where at any time
aggregation processes are able to conglomerate small objects.
Indeed, a study of the time evolution of our mass function (Figure 2)
shows how hierarchical clustering displaces power from  small
to large scales. The presence of a peak in the mass function
confirms the existence of a time--dependent characteristic mass.
This peak, however, becomes less and less prominent
as time goes on.
It is easy to show that the maximum value of the mass function is reached for
$M/M_R \simeq (\alpha / \alpha +1)^{1/2\alpha}
= \left[(n_p+3)/(n_p+9)\right]^{3/(n_p+3)}$.

We can now apply Eq.(\ref {rps}) to a physically sensible model, such as
the standard Cold Dark Matter (CDM) scenario.
We compute the appropriate $\Lambda (M)$ by using 
the transfer function given by Bardeen et al. (1986),
with $\Omega _X =1$ and $H_0= 50$ km s$^{-1}$ Mpc$^{-1}$, 
and by normalizing the spectrum so that
the mass variance is unity in a top--hat sphere of radius $8~h^{-1}$ Mpc.
As Figure 3 shows, also in this case the mass function
presents a strong low--mass cutoff.
In fact, the number density of halos reaches its maximum
at $M\simeq 4.67 \times 10^{11} ~{\rm M}_\odot$ and is strongly damped at
much smaller masses.
We will examine the physical reliability of this cut--off in the following
section. 

To understand how all the mass is divided among the various objects,
in Figure 3
we plot also the ``multiplicity function'' $M^2n(M)/\langle \varrho\rangle$
which gives the mass fraction contained by halos in unit range of $\ln M$.
Obviously, the $d \ln M$ integral of this dimensionless distribution,
performed over the whole mass spectrum, gives unity.

In Figure 4 we show the time evolution of our mass and
multiplicity functions; due to the characteristic scales inherent in the CDM
power--spectrum the self--similarity property is now lost,
even though the mass once again flows from small objects to bigger ones.
In order to quantitatively follow this process,
we study the time dependence of the typical cluster mass of the
distribution as defined in the kinetic theory of aggregation
(see, e.g., van Dongen \& Ernst 1988)
\begin{equation}
\label{mcl}
M_{cl}(z)\equiv{\int_0^\infty M^2 n(M,z) \, dM \over
\int _0^\infty M\, n(M,z)\, dM}
=\int_0^\infty {M^2 n(M,z) \over \langle \varrho \rangle}\, dM.
\end{equation}
In the interval $0\leq z \leq 2$, our results are approximately described
by the power--law
\begin {equation}
\label{evocl}
M_{cl}(z)=M_{cl}(0) (1+z)^{-\beta}
\end{equation}
with $M_{cl}(0)
\simeq 6.39 \times 10^{14} ~{\rm M}_\odot$ and $\beta\simeq 3.36$; however, a
deeper analysis reveals that the quantity
$\beta_{eff}(z)\equiv | {d\ln M_{cl}(z) \over d \ln (1+z)}|$
slowly increases with $z$.
As one would expect, we find $\beta_{eff}(z)\sim {1/\alpha _{eff}(z)}$
where $\alpha _{eff} (z) \equiv -{1\over 2} {d \ln \Lambda (M) \over d \ln M}|
_{M_{cl}(z)}$.
We stress that for the Press--Schechter formula
this equality does not hold because the integral that defines $M_{cl}(z)$
comes out mostly sensitive to low--mass abundances
(Colafrancesco, Lucchin \& Matarrese 1988).

According to high resolution $N$--body simulations, for
$10^{10} ~{\rm M}_\odot \leq M \leq 10^{13} ~{\rm M}_\odot$
the mass function of a standard CDM scenario is well described by a power--law
of index $-2.0\pm 0.1$ in a wide redshift range (Brainerd \& Villumsen 1992).
This feature is not displayed by the Press--Schechter solution whose
logarithmic derivative assumes values $\sim -1.8$ in the
same interval. So, in order to test our solution, we study
the functional dependence of its slope on mass.
We find that only for $z ~\magcir 2$ our solution
behaves like $M^{-2}$, with very good approximation
in the mass interval under consideration.
Anyway, we notice that the choice of the values of the parameters
$\delta _c$  and $\upsilon=M(k_f) k_f^3/ \langle \varrho \rangle $
plays a fundamental role in this kind of comparison. We remind, for example,
that the value $\delta _c=1.686$, obtained from the spherical collapse model,
should not be preferred when one uses a sharp $k$--space filter.
Probably, as suggested by Williams et al. (1991), the wisest choice is to use
numerical simulations to select $\delta_c$ and $\upsilon$ as best fitting
parameters (see also Bond \& Myers 1993, Ma \& Bertschinger 1994,
Klypin et al. 1995, Monaco 1995). It is known however that these best fit
values depend both on the filter used and on the method chosen to 
identify the halos in the simulations (Lacey \& Cole 1994, Gelb \&
Bertschinger 1994). 

\section{Discussion and Conclusions}

In this paper we have derived a new analytical expression for the dark
halos' mass function that develops in a hierarchical clustering scenario.
This result has been achieved by simply modifying the Press--Schechter model
to allow for mass semi--positivity. 

Technically speaking, in every point of configuration space
we have studied the random walk of the coarse--grained density field as a
function of the smoothing scale. Essentially, we have derived
the probability distribution of the filtering lengths that characterize
the events of first up--crossing of a critical threshold $\delta_c$
in the presence of a barrier set in $\delta =\delta_v =-1$.
This boundary had to assure the reflection of the incident ``random--walkers''
to allow for mass semi--positivity.

For power--law spectra we have obtained a mass distribution
whose low--mass tail, for any $z$, has the general form
\begin{equation}
n(M,z)\simeq C_1(z) M^{-2(\alpha+1)}\exp {\left(- C_2(z) M^{-2\alpha} \right)}
\ \ \ \ \ \ \ M\ll M_R(z) ,
\end{equation}
where the characteristic mass $M_R$ and the
parameters $C_1$ and $C_2$ depend on the selected threshold and 
power index.
Furthermore, the mass function presents a maximum for
$M/M_R\simeq \left[(n_p+3)/(n_p+9)\right]^{3/(n_p+3)}$ and
in the high--mass tail
asymptotically reaches the Press--Schechter solution.
During its time evolution the mass multiplicity function keeps
self--similar.

Working directly with many realizations of a Gaussian random field
and performing an object by object analysis,
a number of authors (Bond et al. 1991, Williams et al. 1991, White 1995)
showed that for $M \ll M_*(z)\equiv M_0
/[\delta_c(1+z)]^{1/\alpha}$ the excursion set approach
does not mimic with good approximation
the kinetics of mass aggregation
in a hierarchical scenario, while for $M ~ \magcir M_*$
the correspondence appears quite satisfactory.
Since for many spectral indices, the mass for which our function reach
its maximum value is of the same order of magnitude as $M_*$,
it would be interesting to test our solution against
numerical simulations.

A careful analysis of the mass function of scale--free models in
an Einstein--de Sitter universe has been recently performed by Lacey \& Cole 
(1994), using the high resolution ${\rm P}^3{\rm M}$ code of Efstathiou et al. (1985).
Taking advantage of self--similar scaling, they succeeded in reducing
Poisson fluctuations in counts by averaging the various outcomes obtained 
at different timesteps of the same simulation.
In such a way they could amplify the available mass--range.
Nevertheless, though they considered three different spectral indices 
($n=-2,-1,0$), only for $n=0$ their low--mass limit is such as to make
potentially observable the cut--off implied by our solution.
Moreover, $n=0$ is the only index for which the different choice of the
filtering method (top--hat with $M(R_f)=4 \pi \langle \varrho \rangle R_f^3/3$
in the work of Lacey and Cole, sharp--k with  $\upsilon = 6 \pi^2$ in ours)
has no effect. However, the numerical results agree quite well
with the Press--Schechter solution showing no trace of a low--mass cut--off.
For example, for $n=0$ and $M/M_R \simeq 0.2$ (which is the lowest mass
considered in the mass function obtained from the numerical data)
our solution is almost a factor of 
3 smaller than the Press--Schechter one which, on the contrary, overestimates 
the counts by $\sim 30 \%$. 

We are not surprised by this disagreement, since the method we used
to allow for mass semi--positivity is a very crude one. We think that only by
dealing with a field that has the correct statistical properties one can 
determine the exact position of the cut--off. Our algorithm is only able to 
show that a low--mass cut--off must exist and to explain the origin of 
the divergence of the Press--Schechter solution for $M \to 0$, as being due to 
unphysical negative mass events in the Gaussian distribution. 

In order to test our solution also in the CDM case we refer to a recent
paper by Efstathiou (1995), who 
analysed the redshift evolution (for  $1 \leq z \leq 3$
and $M > 10 ^{10} {\rm M}_\odot$) of the integrated mass function
$N(>M,z)=\int _M^\infty n(M^{\prime},z) dM^{\prime}$ in a highly biased,
CDM dominated
Einstein--de Sitter universe.
By comparing our results with these data we find a fairly good agreement
for $z=3$, while we predict a larger number of counts ($\sim 75 \%$ more)
for $M\simeq 10^{10}{\rm M}_\odot$ at $z=1$.
Actually, at high $z$ we are testing the high--mass end of the distribution,
where our solution is practically indistinguishable from the one of
Press and Schechter, that fits the data quite well.
On the contrary, at intermediate redshifts, the mass--range in analysis
involves masses just above the peak, 
where our solution implies more objects than the Press--Schechter one 
(see also the figures 3 and 4).

Even though we faced the negative--mass problem in quite a formal way, our 
treatment allowed to understand the origin of the ``infrared divergence" of the
Press--Schechter theory. We do not claim that we are able to indicate the
best solution, since we believe that the correct answer
must rely on the intrinsic non--Gaussian nature of any density
fluctuation field.
In any case, we think that the different behaviour of our solution compared
to the Press--Schechter one, if confirmed and improved by more detailed 
modeling, might be useful to develop a consistent picture for the 
formation of dark halos. 
In fact, it is known that, assuming a constant mass--to--light ratio,
the Press--Schechter formula predicts too many low--mass objects with
respect to the observed galaxy luminosity function (Bond et al. 1991).
However, to deal with galaxies one should consider many
astrophysical and hydrodynamical effects (gas cooling in non
stationary conditions, star formation and so on) that are supposed to be
fundamental issues of galaxy formation but are very hard to model.
These subjects are clearly beyond the purposes of this work.

In summary, we have shown that the low--mass divergence of the
Press--Schechter mass function can be ascribed to the use of
Gaussian fields to describe the cosmological density fluctuations, which 
assign a finite probability to events with a
negative mass; since this probability comes out directly proportional
to the variance, in a hierarchical clustering model the
reliability of theoretical predictions should get worse as the variance
increases i.e. as the mass decreases. We have shown that this is indeed the
case. We believe that a reliable model able to make quantitative predictions
should not leave truly non--Gaussian fields out of consideration.
Even though there are good reasons to think that the primordial
gravitational potential was very nearly Gaussian distributed, one should derive
the statistical features of $\delta$ from the non--linear
fluid--dynamical equations. Only in this way one would be able
to obtain the correct statistical properties of the density fluctuation field.
We will return to this subject in a future work.

\section* {Acknowledgments} The Italian MURST is acknowledged for financial
support.

\appendix
\section{Void regions and the mass function with two absorbing barriers}

In Section 3, by modifying the excursion--set approach, 
we have seen how the introduction of a second barrier, able to
limit the dispersion of the values assumed by $\delta({\bf x},R_f)$,
would produce relevant changes in the behaviour of the mass function.
We want here to investigate the effect of changing the nature
of this new barrier, with the aim of simulating a particular physical
situation, namely the existence of void regions. 

Let us consider a physical density fluctuation field, which is clearly
larger than or equal to
$-1$ everywhere in space; provided the applied filter is also
semi--positive definite and correctly normalized, the only possibility to
obtain the value $-1$ for the smoothed density fluctuation field is
realized when the entire region which contributes to the convolution integral
defining $\delta({\bf x},R_f)$ is void.

Let us then imagine to use a filter with radius $\hat R_f$
and find $\delta (\hat {\bf x},\hat R_f)=-1$ in some arbitrary point
$\hat {\bf x}$; for every smoothing radius smaller than $\hat R_f$
(therefore for every variance $\Lambda>\Lambda(\hat R_f)$) the
considered region must still be void. In practice one must obtain
$\delta(\hat {\bf x},\Lambda)=-1$ for every $\Lambda>\Lambda(\hat R_f)$.
Therefore, once the value $-1$ has been attained at the variance
$\Lambda (\hat R_f)$, the density field in the point
$\hat {\bf x}$ cannot assume any other value:
also the barrier set at $\delta =\delta_v$ behaves as an absorbing one.
In such a way one succeeds in accounting for all the points included
inside void regions.
For each of these points $\hat {\bf x}_v$, in fact,
there exists a smoothing radius $\widetilde R_f$ corresponding to
the minimum distance of the point to the boundary of the void region,
such that, for any $R_f<\widetilde R_f$, one measures
$\delta({\bf x}_v,R_f)=-1$
(obviously, we are dealing with filter functions that do not vanish
only in a finite region of space, as, for example, the top--hat one). 

Once again, in order to obtain quantitative results, we need to use
the sharp $k$--space filter even though this weakly violates our
hypotheses. Then, one has to work out
the Fokker--Planck equation (\ref{fp})
with the boundary conditions ${\cal W}(\delta_v,\Lambda)=0$,
${\cal W}(\delta_c,\Lambda)=0$ and the initial condition
${\cal W}(\delta ,0)=\delta_D(\delta)$.
Hence, using the same notation of section 3, one has to solve
Eq.(\ref{fpx}) with the boundary conditions ${\cal W}(0,t)=0$,
${\cal W}(A_v+A_c,t)=0$ and the initial condition
${\cal W}(x,0)=\delta_D(x-A_v)$
(the previous parameter $R_v$ has been replaced by $A_v$ to emphasize the
different nature of the barrier set in $\delta_v$).

Once again one can proceed by separation of variables, finding
${\cal W}(x,t) = 2 \sum_{n=1}^\infty \phi_n(x)
 \phi_n(A_v) \exp(-{1\over 2}
\lambda_n t)$, with $\lambda_n = [n\pi/(A_v+A_c)]^2$ and
$\phi_n(x)= (A_v+A_c)^{-1/2} \sin\left( \sqrt{\lambda_n}x\right)$.
One then gets
\begin{equation}
{\cal W}(x,t)={2\over A_v+A_c} \sum_{n=1}^\infty
\sin\left({n\pi\over A_v+A_c}A_v\right) \sin\left({n\pi\over A_v+A_c}x\right)
\exp\left(-{n^2\pi^2\over 2(A_v+A_c)^2}t\right),
\end{equation}
so that one can easily compute the first--crossing rates ${\cal T}_v$ and
${\cal T}_c$
across the barriers respectively set in $x=A_v$ and $x=A_c$; one obtains
\begin{equation}
{\cal T}_v(t) = - {\cal J}(0,t) =
{\pi\over (A_v+A_c)^2} \sum_{n=1}^\infty
n \sin\left({n\pi\over A_v+A_c}A_v\right)
\exp\left(-{n^2\pi^2\over 2(A_v+A_c)^2}t\right)
\end{equation}
and
\begin{equation}
{\cal T}_c(t) = {\cal J}(A_v+A_c,t) =
{\pi\over (A_v+A_c)^2} \sum_{n=1}^\infty
(-1)^{n+1} n \sin\left({n\pi\over A_v+A_c}A_v\right)
\exp\left(-{n^2\pi^2\over 2(A_v+A_c)^2}t\right)
\end{equation}
respectively. To compute the mass function one only needs
$f(\Lambda,\delta_c(z))\equiv {\cal T}_c(\Lambda)$, which replaced into
Eq.(\ref{genps}) gives
\begin{equation}
\label{aps1}
n(M,z) =  {\pi \langle \varrho \rangle \over (1+\delta_c)^2 (1+z)^2}
{\Lambda (M)\over M^2} \left| {d \ln \Lambda \over d\ln M}
\right|
\sum_{n=1}^\infty (-1)^{n+1} n \sin \left({n\pi \over 1+\delta_c}
\right) \exp\left( - {n^2 \pi^2 \Lambda (M)\over
2 (1+\delta_c)^2 (1+z)^2}\right) .
\end{equation}

However,
before going on we need to face a new situation.
Both in the case of a single absorbing barrier
set at $\delta=\delta_c$ and in the case of a second reflecting boundary placed
at $\delta=\delta_v$, all the ``random--walkers" are eventually going to cross
the barrier set at $\delta =\delta_c$; on the contrary, in the case
where the second barrier is also an absorbing one,
a relevant fraction of all realizations crosses the boundary
set at $\delta =\delta_v$,
thereby giving no contribution to collapsed objects.
In order to quantify this fraction we need to compute
the two quantities
${\cal P}_v$ and ${\cal P}_c$
representing the crossing probabilities respectively at
$\delta =\delta_v$ and $\delta =\delta_c$. One has
\begin{equation}
{\cal P}_v = \int_0^\infty {\cal T}_v(t)\, dt={2\over \pi} \sum _{n=1}^\infty
{1\over n} \sin \left( {n\pi A_v \over A_v+A_c} \right)
\end{equation}
and
\begin{equation}
{\cal P}_c = \int_0^\infty {\cal T}_c(t)\, dt={2\over \pi} \sum _{n=1}^\infty
{(-1)^{n+1}\over n} \sin \left( {n\pi A_v \over A_v+A_c} \right).
\end{equation}
To compute the series one has
to Fourier expand the following elementary functions defined in the interval
$(0,2L)$
\begin{equation}
f(x)={x \over 2L}={1\over 2}-{1\over \pi}\sum _{n=1}^\infty
{1\over n} \sin {n\pi x \over L} ,
\end{equation}
\begin{equation}
g(x)=\cases {\displaystyle {x \over L} & if $0<x<L$\cr
\noalign {\vskip 10pt}
\displaystyle {x\over L}-2& if $L<x<2L$\cr }
={2\over \pi}\sum _{n=1}^\infty
{(-1)^{n+1}\over n} \sin {n\pi x \over L} ,
\end{equation}
from which one obtains ${\cal P}_c=A_v /(A_v+A_c)$ and
${\cal P}_v = A_c /(A_v+A_c)$.
One can conclude that, for the cosmologically relevant cases, one has
${\cal P}_v = \delta_c /(\delta_v+\delta_c)=\delta _c/(1+\delta _c)$
at any redshift.

We can interpret this result in two different ways:
{\it i)} the particles that, because of their peculiar initial conditions, 
end up their existence in ``voids'' do not form halos at all and then only a 
fraction ${\cal P}_c = \delta_v /(\delta_v+\delta_c)=1/(1+\delta _c)$ of the 
total mass is collapsed in bound object;
{\it ii)} the regions that form ``voids'' are really depleted of mass
(this would be obviously true if we were following the real dynamics
of the field until
it reaches the value $-1$) and then the mass that initially was contained
in them should have flowed into the overdense regions.
The mass function associated with the former interpretation is given in
Eq. (\ref {aps1}): one of its main properties is that it asymptotically
recovers the Press--Schechter form at the high--mass end.
However, the latter interpretation is the only one 
that can assure that all the mass
is collapsed in object as it is usually required in hierarchical models.
In fact, if one wants $\int _0^\infty M\, n(M) dM = \langle \varrho \rangle $
to hold, one needs to modify Eq.(\ref {aps1})
and more generally Eq. (\ref {genps}) by replacing 
$\langle \varrho \rangle$ with the mean density of non--empty regions
$\langle \varrho _M \rangle=\langle \varrho \rangle / {\cal P}_c=
(1+\delta _c)\langle \varrho \rangle$.

By introducing the parameter $f_{norm}$ 
($0\leq f_{norm} \leq \delta _c$) indicating 
the fraction of mass flowed from underdense into overdense regions we can write

\begin{equation}
\label{aps}
n(M,z) = {(1+f_{norm})\pi \langle \varrho \rangle \over (1+\delta_c)^2 (1+z)^2}
{\Lambda (M) \over M^2} \left| {d \ln \Lambda \over d\ln M}
\right|
\sum_{n=1}^\infty (-1)^{n+1} n \sin \left({n\pi \over 1+\delta_c}
\right) \exp\left( - {n^2 \pi^2 \Lambda (M)\over
2 (1+\delta_c)^2 (1+z)^2}\right) .
\end{equation}

For scale--free spectra the mass function then becomes
\begin{equation}
\label{ma}
n(M,z)={4 \alpha (1+f_{norm}) \over \pi} {\langle \varrho \rangle \over M^2}
\left({M\over M_A(z)}\right) ^{-2\alpha}
\sum_{n=1}^\infty (-1)^{n+1} n \sin \left({n\pi \over 1+\delta_c}
\right) \exp\left( - n^2
 \left({M\over M_A(z)}\right) ^{-2\alpha}\right) ,
\end{equation}
where $M_A(z)\equiv 2^{1/\alpha} M_R(z)$.

In Figure A1 this solution is compared with the Press--Schechter one:
qualitatively the new mass function looks like that of Eq.(\ref{mr}).
Once again the multiplicity function evolves in a self--similar way; 
for $n_p >-2$ 
the maximum value of the mass function is reached for $M/M_A\simeq
[(\alpha / \alpha +1)]^{1/2\alpha}$.
This implies that this distribution cuts off at a larger mass compared 
with the solution given in the text.

This feature is present also when one considers the standard CDM
power--spectrum (Figure A2);
in this case the peak of the mass function is reached for
$M=5.77 \times 10^{13} ~{\rm M}_\odot$ while the typical cluster mass
comes out at $M_{cl}(0)=1.48\times 10^{15} ~{\rm M}_\odot$.
The time evolution of our new mass function
is consistent with Eq.(\ref{evocl}), where $\beta \simeq 3.27$.
Once again $\beta_{eff}(z)$ and $\alpha_{eff}(z)$ turn out to be
tightly correlated: their product is approximately equal to one.
This agreement is remarkable for $z\simeq 0$ while it gets worse
as $z$ increases. This is exactly the behaviour we would expect
since, as $z$ grows, the integral that defines $M_{cl}(z)$
takes contributions from a wider mass interval. 

\newpage

\section* {Figure Captions}

{\bf Figure 1}
The mass function $n(M)$ obtained by allowing
for mass semi--positivity 
(solid line) is 
 compared, for different scale--invariant
spectra, with the Press--Schechter solution (dotted line).
The behaviour of the dimensionless and time--independent
distribution $M_R^2 n(M)/\langle \varrho \rangle$ is shown as a function of 
$M/M_R$, where $M_R$ is a redshift--dependent 
characteristic mass defined in the text.

\medskip
\noindent 
{\bf Figure 2}
Time evolution of $M_0^2 n(M) /\langle \varrho \rangle$ for
a power--law spectrum with $n_p=1$.

\medskip
\noindent
{\bf Figure 3}
The present--day mass function obtained by accounting
for the mass semi--positivity constraint and the related multiplicity
function $M^2 n(M) /\langle \varrho \rangle$ (solid lines)
are compared with
their Press--Schechter counterparts (dotted lines) for a standard CDM scenario.
The power--spectrum is obtained starting from a primordial power--law
with $n_p=1$ and using the transfer function given by Bardeen et al.
(1986) with the choices $\Omega_X =1$, $h=0.5$, $\sigma_8 =1$,
$\delta_c =1.686$. 

\medskip
\noindent
{\bf Figure 4}
The mass and the multiplicity functions
for the standard CDM model described in Figure 3
are shown at three different redshifts.

\medskip
\noindent
{\bf Figure A1}
The time--independent function $M_A^2 n(M)/\langle \varrho \rangle$, 
obtained
by allowing for
the existence of void regions and by imposing $f_{norm}=\delta _c$
(solid line), is compared, for different
scale--invariant spectra, with the Press--Schechter solution (dotted line).

\medskip
\noindent
{\bf Figure A2}
Present mass and multiplicity functions in a standard CDM scenario.
The solutions obtained with the absorbing barrier at $\delta_v$ 
and with $f_{norm}=\delta _c$
are represented by a solid line, while the Press--Schechter ones
are plotted with a dotted line.

\newpage

\begin{figure}
\centerline {\input fig1e.tex \input fig1d.tex}
\vskip 1truecm
%\centerline {\input fig1c.tex}
%\vskip 0.5truecm
\centerline {\input fig1c.tex \input fig1b.tex}
\label{rspot}
\end{figure}

\vskip 3truecm
\centerline {\bf Figure 1}

\newpage

\vskip 8truecm

\begin{figure}
\centerline{\input cri2.tex }
\label{erisp}
\end{figure}

\vskip 3truecm
\centerline {\bf Figure 2}
\newpage

\vskip 8truecm
\begin{figure} 
\centerline {\input cri3se.tex \input tic3b.tex }
\label{cdmri}
\end{figure}

\vskip 3truecm
\centerline {\bf Figure 3}
\newpage

\vskip 8truecm
\begin{figure} 
\centerline{\input cri4.tex \input tic4b.tex}
\label{rcdmev}
\end{figure} 

\vskip 3truecm
\centerline {\bf Figure 4}
\newpage

\vskip 4truecm
\begin{figure}
%\centerline {\input fig5e.tex \input fig5d.tex}
%\vskip 1truecm
%\centerline {\input fig5c.tex}
%\vskip 0.5truecm
\centerline {\input fig5c.tex \input fig5b.tex}
\label{aspot}
\end{figure}

\vskip 3truecm
\centerline{\bf Figure A1}
\newpage

\vskip 8truecm
\begin{figure} 
\centerline {\input cri7se.tex \input fig7b.tex}
\label{cdmas}
\end{figure}

\vskip 3truecm
\centerline {\bf Figure A2}
\newpage


\begin{thebibliography}{}

\bibitem [\protect\refname{Bardeen et al. }1986]{bbks}
Bardeen J.M., Bond J.R., Kaiser N., Szalay A.S., 1986, ApJ, 304, 15
\bibitem [\protect\refname{Bond et al. }1991]{bcek}
Bond J.R., Cole S., Efstathiou G., Kaiser N., 1991, ApJ, 379, 440
\bibitem [\protect\refname{Bond \& Myers }1994]{bm94}
Bond J.R., Myers S.T., 1993, CITA preprint 93/27
\bibitem [\protect\refname{Brainerd \& Villumsem }1992]{brvi}
Brainerd T.G., Villumsen J.V., 1992, ApJ, 394, 409
\bibitem [\protect\refname{Catelan et al. }1994]{ccmm}
Catelan P., Coles P., Matarrese S., Moscardini L., 1994, MNRAS, 268, 966
%\bibitem [\protect\refname{Cavaliere \& Menci}1993]{cm93}
%Cavaliere A., Menci N., 1993, ApJ, 407, L9
\bibitem [\protect\refname{Cavaliere et al. }1995]{cmt95}
Cavaliere A., Menci N., Tozzi P., 1995, ApJ, in press
\bibitem [\protect\refname{Chandrasekhar }1943]{chan}
Chandrasekhar S., 1943, Rev. Mod. Phys., 15, 1
\bibitem [\protect\refname{Colafrancesco et al. }1989]{cola}
Colafrancesco S., Lucchin F., Matarrese S., 1989, ApJ, 345, 3
\bibitem [\protect\refname{Cole }1991]{cole91}
Cole S., 1991, ApJ, 367, 45
\bibitem [\protect\refname{Efstathiou }1995]{ef95}
Efstathiou G., 1995, MNRAS, 272, L25
\bibitem [\protect\refname{Efstathiou et al. }1985]{e85}
Efstathiou G., Davis M., Frenk C.S, White S.D.M., 1985, ApJS, 57, 241
\bibitem [\protect\refname{Efstathiou et al. }1988]{e88}
Efstathiou G., Frenk C.S, White S.D.M., Davis M., 1988, MNRAS, 235, 715
\bibitem [\protect\refname{Gelb & Bertschinger} 1994]{geber94}
Gelb J.M., Bertschinger E., 1994, ApJ, 436, 467
\bibitem [\protect\refname{Klypin et al. }1995]{kly95}
Klypin A., Borgani S., Holtzman J., Primack J.R., 1995, ApJ, 444, 1
\bibitem [\protect\refname{Lacey & Cole } 1993]{lc93}
Lacey C., Cole S., 1993, MNRAS, 262, 627
\bibitem [\protect\refname{Lacey & Cole } 1994]{lc94}
Lacey C., Cole S., 1994, MNRAS, 271, 676
\bibitem [\protect\refname{Lilje } 1992]{lilje}
Lilje P. B., 1992, ApJ, 386, L33
\bibitem [\protect\refname{Lucchin \& Matarrese}1988]{lm88}
Lucchin F., Matarrese S., 1988, ApJ, 330, 535
\bibitem [\protect\refname{Ma \& Bertschinger }1994]{maber94}
Ma C.--P., Bertschinger E., 1994, ApJ, 434, L5
\bibitem [\protect\refname{Manrique \& Salvador--Sol\'e}1995]{mass95}
Manrique A., Salvador--Sol\'e E., 1995, ApJ, 453, 6
\bibitem [\protect\refname{Monaco }1995]{mon95}
Monaco P., 1995, ApJ, 447, 23
\bibitem [\protect\refname{Peacock \& Heavens }1990]{ph90}
Peacock J.A., Heavens A.F., 1990, MNRAS, 243, 133
\bibitem [\protect\refname{Press \& Schechter }1974]{ps74}
Press W.H., Schechter P., 1974, ApJ, 187, 425
\bibitem [\protect\refname{Risken }1989]{risken}
Risken H., 1989, The Fokker--Planck Equation, Springer--Verlag, Berlin
\bibitem [\protect\refname{Sheth}1995]{sh95}
Sheth R.K., 1995, MNRAS, 274, 213
\bibitem [\protect\refname{van Dongen \& Ernst}1988]{VDE}
van Dongen P.G.J., Ernst M.H., 1988, J. Stat. Phys., 50, 295
\bibitem [\protect\refname{White} 1995]{whi95}
White S.D.M., 1995, in Schaeffer R., ed., Les Houches Lectures, in press  
\bibitem [\protect\refname{Williams et al. }1991]{whps}
Williams B.G., Heavens A.F., Peacock J.A., Shandarin S.F., 1991,
MNRAS, 250, 458

\end{thebibliography}
\end{document}